# Numerical study of flow physics in supersonic base-flow with mass bleed

Pratik Das, Ashoke De[*]

Department of Aerospace Engineering, Indian Institute of Technology, Kanpur, India-208016

*Corresponding author: Ph:+91-5122597863, Fax: +91-5122597561
E-mail: ashoke@iitk.ac.in

**ABSTRACT**

Large Eddy Simulation (LES) with dynamic sub-grid scale eddy viscosity model has been applied to numerically investigate the evolution of complicated flow structures in supersonic base flow with mass bleed. Mean flow properties obtained from numerical simulations, such as axial velocity, pressure on the base surface, have been compared with the experimental measurements to show that LES is able to predict the mean flow properties with acceptable accuracy. The data obtained from LES has been further analyzed to understand the evolution of coherent structures in the flow field. Periodical shedding of vortical structures from the outer shear layer has been observed and it has also been found that this vortex shedding is associated with the flapping of the outer shear layer. The frequency of flapping of the outer shear layer has been found out and the phase-averaged streamlines have been analyzed to further study the evolution of vortical structures associated with this flapping. The phase-averaged streamline plots clearly elucidate the evolution of vortical structures along the outer shear layer. Further, the study of these structures is investigated by performing Proper Orthogonal Decomposition (POD) analysis of the data, obtained along the central plane in the wake region. The POD results also seem to agree well with the observations made in the phase averaged streamline plots, as the concentrated energy and enstropy are observed in the outer shear layer with fewer POD modes.

## 1. INTRODUCTION

The aerodynamic performance of bullets, projectiles, and ballistic missiles are often compromised as they suffer massive pressure drag while traveling at supersonic speed. Due to design constraints, the design of this kind of object often features a blunt base with sharp corner similar to a cylindrical afterbody. As the turbulent boundary layer separates at the sharp corner, a low velocity, and a low-pressure recirculation region is formed which is separated from the supersonic flow outside by the compressible shear layer. This situation leads to the formation of partial vacuum just behind the base surface, thereby leading to massive pressure drag, otherwise known as base drag. Over the years, different passive techniques like boat-tailing, base cavity as well as active techniques like base burning, base bleeding etc. have been developed to reduce the base drag; yet the complicated fluid dynamics of the problem has always eluded practicing engineers and scientists from devising optimal parameters for application of these techniques. To obtain the optimized operating parameters for these techniques, it is crucial that the pressure on the base surface is predicted accurately as the key parameters are changed.

In this study, the complicated fluid dynamics associated with the base bleed technique has been numerically investigated. In this technique, a subsonic jet is injected from the base into the wake region of the cylindrical afterbody. As the low momentum fluid in the recirculation region gains momentum due to the mixing of higher momentum fluid from the jet, the recirculation region is pushed further downstream, away from the base surface and sudden rise in pressure on the base surface is also observed. How far the recirculation region is being pushed would depend on the mass flow rate of the bleeder jet, which also affects the average pressure on the base surface. The mass flow rate of the bleeder jet is quantified using a non-dimensional injection parameter (*I*), defined as the ratio of mass flow of the subsonic jet and the product of base area and mass flux of the free stream flow outside the recirculation region (Eq. 1). Though the injection parameter does not take the momentum of the jet or the incoming boundary layer thickness on the after-body into account, in several previous experimental studies [1-3] on base bleed, it has been observed that the base pressure ratio has a strong dependence on the injection parameter. In mid-nineties, Herrin and Dutton [1] performed a detailed experimental study of supersonic base flow, followed by the experiments studies performed by Mathur and Dutton [2-3]. They studied the effects of subsonic mass injection in the wake from the base on the base pressure. They have performed an extensive study of flow properties in the near wake region behind a cylindrical afterbody of 63.5 mm diameter with a bleed orifice of diameter 25.4mm, in perfect axial alignment with a Mach 2.46 flow (approach Mach number). Their experimental facility was specifically designed to maintain the axial alignment while reducing the effects of support strings on the flow field. The experiments performed by Mathur and Dutton [2-3] provided an excellent database for validating the results obtained from CFD solvers and assessing the performance of different numerical techniques and mathematical models applied to solve this complicated problem. In their experiments, Mathur and Dutton [2-3] observed that the maximum base pressure ratio can be achieved at the injection parameter value of I= 0.0148.

In the past couple of decades several numerical studies of supersonic base flow have been performed [4-17], while only a few numerical studies of base bleed [18-22] were performed mostly focused on validating the experimental results obtained by Mathur and Dutton [2-3]. Sahu and Heavy [23] performed RANS (Reynolds Averaged Navier Stokes) simulation with limited success in predicting the distribution of axial velocity component along the centerline. They have also shown that the two equation *k-ε* turbulence model performed better than the Baldwin-Lomax algebraic model. Lee et al. [24] performed a similar RANS study while using the standard k-ω turbulence model in the Fluent solver. Bournot et al. [25] performed a numerical study of the base bleed and showed that the introduction of reactive particles in the subsonic bleeder jet may improve effectiveness by leading to higher pressure on the base surface. Shin and Choi [26] obtained good agreement with the experimental results when they performed DDES (Delayed Detached Eddy Simulation) study of the supersonic base flow and effect of the base bleed technique.

It is quite evident from the available literature that most of the previous numerical works were primarily focused on the predictions of mean flow field successfully, whereas little effort has been made to identify the evolution of flow structures in the near wake region of the base. Study of coherent flow structures in the supersonic regime is still a sparsely explored area of research. Current research is aimed towards a comprehensive study of flow structures that appear when a subsonic jet is injected from the base into the wake region of a cylindrical afterbody placed in supersonic flow. In the present study, two injection parameter values, i.e. I=0.0148 and I=0.0226 from the base bleed experiments of Mathur and Dutton [2-3], have been chosen as the test cases. LES of both the cases has been performed to resolve the evolution of large-scale structures in the wake region. POD in the wake region has also been carried out to successfully predict the most energetic flow structures present in the flow field. We have also compared our LES results with the experimental data obtained by Herrin & Dutton [1] for supersonic base flow without any flow control. Detailed analysis of our numerical study of base flow without flow control has already been reported and can be found in the published literature [27].

$$Injection\ Parameter(I) = \frac{\dot{m}_{bleed}}{\pi R_{base}^2 \times \rho_{inf} \times U_{inf}} \quad [1]$$

$$T_{bleed} = T_{0,bleed} - \frac{\gamma - 1}{R\gamma} \times \frac{U_{bleed}^2}{2} \quad [2]$$

$$P_{bleed} = IRT_{bleed} \frac{A_{base} \times \rho_{inf} \times U_{inf}}{A_{bleed} \times U_{bleed}} \quad [3]$$

## 2. NUMERICAL DETAILS

The Favre-filtered conservation equations for mass, momentum, energy and species transport are solved in the present work as used by the authors in the previous work [27-32]. To model the turbulent eddy viscosity ($v_t$), LES is used so that the energetic larger-scale motions are resolved, and only the small-scale fluctuations are modeled. The sub-grid stress modeling is done using a dynamic Smagorinsky model [33-34]. For POD analysis, we have used the method of snapshots as detailed out in the literature [27, 35-40] while considering the effect of compressibility through temperature correction. Usually, the aim of POD is to find out the set of orthonormal basis vectors of an ensemble of data in a lower dimension space to identify the most predominant structures in the data which are often hidden in the ensemble of the data. More details regarding the flow solver, numerical schemes, POD can be found in the literature [27-40].

### 2.1 Flow modeling using LES

To model the turbulence, LES is used where the large scale structures are resolved and the small scale structures are modeled. Hence, the Favre-filtered governing equations for the conservation of mass, momentum, energy and species transport are solved in the present work [27-32].

Dynamic Smagorinsky model is used for sub-grid stress modeling [33-34], where the gradient approximation is invoked to relate the unresolved stresses to resolved velocity field and given as:

$$\widetilde{u_i u_j} - \tilde{u}_i \tilde{u}_j = -2\nu_t \bar{S}_{ij} \quad [4]$$

$$\text{Where } \nu_t = C_s^2 (\Delta)^2 |\bar{S}| \quad [5]$$

$$\bar{S}_{ij} = \left( \frac{\partial \bar{u}_i}{\partial x_j} + \frac{\partial \bar{u}_j}{\partial x_i} \right) \quad [6]$$

$$|\bar{S}| = \sqrt{2\bar{S}_{ik}\bar{S}_{ik}} \quad [7]$$

And S is the mean rate of strain. The coefficient $C_s$ is evaluated dynamically [33-34]. More details regarding the governing equations can be found in the literature [27-32].

**2.2 Numerical Scheme**

A density based, fully coupled FVM based solver has been used to solve the governing equations. A second order Low Diffusion Flux Splitting Scheme has been used to discretize the convective terms (Edwards [41]). All other spacial terms (i.e. diffusion terms) in the governing equations are discretized using second order central difference scheme, while the second order implicit temporal discretization is used. Moreover, the Low Mach number preconditioning (Weiss and Smith [42]) is used to effectively capture the different flow regime in the domain. The parallel processing is done using Message Passing Interface (MPI) technique. More details can be found in the literature [27-32].

**2.2 Mathematical formulation used for POD analysis:**

The objective of POD technique is to find a set of orthonormal basis vectors of an ensemble of data in a lower dimension space such that every member of the ensemble can be decomposed relative to the orthonormal basis while minimizing the error between the ensemble and its projection on the new lower dimension space. POD technique serves two purposes: firstly, it performs reduction by projecting the higher dimensional data into a lower dimensional space and secondly, it reveals the most predominant structures in the data which are often hidden in the ensemble of the data. Use of POD technique in the context of turbulent flows was introduced by Lumley [37]. The mathematical foundation for the application of POD is being briefly presented here in this section.

Let us consider an ensemble of data $U^k$ of dimension 'm' which belongs to Hilbert space $H$, i.e. $U^k \in H | k = 1,2,3,......,m$. Now let us consider another subspace 'S' with dimension $n<m$ and let orthogonal projection onto the subspace $S$ be defined as operator $T_s$. Our aim is to maximize $\overline{\|T_s U^k\|}$ or minimize the error $\overline{\|U^k - T_s U^k\|}$ (since $\|U^k\|^2 = \|U^k - T_s U^k\|^2 + \|T_s U^k\|^2$) where $\|\cdot\|$ defines the induced norm on H and $\overline{\phantom{x}}$ defines averaging over $k$. Let $U^k$ be an ensemble of vector quantity obtained from numerical simulation or experiment such that $U^k = U(x, t^k)$, where $t^k$ is $k^{th}$ time step. From Lumley's [37] basic formulation for calculation of bases spanning subspace $S$ such that the error $\overline{\|U^k - T_s U^k\|}$ is minimized, we arrive at the following eigen value problem:

$$\int_\Omega R(x,x')\emptyset(x')dx' = \lambda \emptyset(x) \qquad [8]$$

Where, $R(x, x')$ is the auto correlation tensor given by,

$$R(x,x') = \overline{U(x,t)U(x',t)} \qquad [9]$$

Equation (8) gives rise to countable infinite eigen values $\{\lambda\}_1^\infty$ and corresponding eigen functions $\{\emptyset\}_1^\infty$. Once these eigen values and eigen functions are known, these can be used to decompose the vector field as

$$U(x,t) = \sum_{n=1}^{\infty} a^n(t)\, \emptyset^n(x) \quad [10]$$

Where, the temporal coefficients, $a^n(t)$, are given as

$$a^n(t) = (U, \emptyset^n) \quad [11]$$

In this study, the method of snapshots proposed by Sirovich [38] has been used to solve the eigen value problem.

Let us consider N number of data sets corresponding to $N$ snapshots in time. Now $U(x,t)$ be a fluctuating vector field with three components $U,V,W$. $U(x,t)$ has a finite value everywhere in the domain and it is square integrable over the domain of interest. Assuming the ensemble is sufficiently large, the auto correlation tensor $R(x, x')$ can be approximated as the following

$$R(x, x') = \frac{1}{N} \sum_{n=1}^{N} U(x, t^n) U^T(x', t^n) \quad [12]$$

Let us assume the basis mode can be written in terms of original dataset as given below,

$$\emptyset(x) = \sum_{n=1}^{N} A(t^n)\, U(x, t^n) \quad [13]$$

Substitution of $R(x, x')$ & $\emptyset(x)$ in equation (8) followed by some rearrangements yield

$$\sum_{n=1}^{N} \left( \frac{1}{N} \int_\Omega U^T(x', t^n) U(x', t^n) dx' \right) A(t^n) = \lambda A(t^n) \quad [14]$$

Now let us define,

$$C = C(i,j) = \frac{1}{N} U^T(x, t^i) U(x, t^j) \quad i,j = 1,2,3,\ldots\ldots\ldots, N \qquad [15]$$

$$A = A^n = A(t^n) \quad n = 1,2,3,\ldots\ldots\ldots, N \qquad [16]$$

Now, equation (14) can be written as,

$$CA = \lambda A \qquad [17]$$

By solving the eigen value problem presented by equation (17) $N$ mutually orthogonal eigen vectors $A_i, i = 1,2,3,\ldots\ldots\ldots, N$ are obtained

Normalized POD modes are given by,

$$\emptyset_i(x) = \frac{\sum_{n=1}^{N} A_i(t^n) U(x, t^n)}{\|\sum_{n=1}^{N} A_i(t^n) U(x, t^n)\|} \qquad [18]$$

In the present study, enstropy and energy based POD analyses on a 2D plane along the centerline have been performed. For enstropy based POD, the fluctuating components of vorticity vector are considered as the components of the auto-correlation matrix.

$$U = [\omega'_x \quad \omega'_y \quad \omega'_z]^T \qquad [19]$$

And the inner product of the field is given by the following equation,

$$\left( U(x, t^i), U(x, t^j) \right) = \int_\Omega \{\omega'_x(x, t^i) \cdot \omega'_x(x, t^j) + \omega'_y(x, t^i) \cdot \omega'_y(x, t^j) + \omega'_z(x, t^i) \cdot \omega'_z(x, t^j)\} dx \qquad [20]$$

To perform energy based POD for compressible flows, it is imperative that the temperature field is also being taken into account along with the velocity components while forming the auto-correlation matrix, as the internal energy field has significant contributions in the total energy field and considerable variation of internal energy is often observed all over the flow domain which is being considered for the POD analysis. So the fluctuation of temperature has also been taken into account along with the fluctuating velocity components.

$$\boldsymbol{U} = [u' \quad v' \quad w' \quad T']^T \quad [21]$$

While the inner product of the field is defined by the following equation,

$$\left(\boldsymbol{U}(x,t^i), \boldsymbol{U}(x,t^j)\right) = \int_\Omega \{u'(x,t^i) \cdot u'(x,t^j) + v'(x,t^i) \cdot v'(x,t^j) + w'(x,t^i) \cdot w'(x,t^j) + \gamma \cdot T'(x,t^i) \cdot T'(x,t^j)\} dx \quad [22]$$

Here $\gamma$ is a scaling factor used to balance the velocity and temperature fluctuation energies. The optimal value of $\gamma$, as suggested by Lumley and Poje [39] and Qamar and Sanghi [40], is given by the following equation.

$$\gamma = \frac{\overline{\int_\Omega \{\rho'^{(x,t^i)} \cdot \rho'^{(x,t^i)} + u'(x,t^i) \cdot u'(x,t^i) + v'(x,t^i) \cdot v'(x,t^i) + w'(x,t^i) \cdot w'(x,t^i)\} dx}}{\overline{\int_\Omega \{T'(x,t^i) \cdot T'(x,t^i)\} dx}} \quad [23]$$

Where '‾' denotes averaging over time.

## 2.1 Boundary Conditions

In this work, two different bleed ratios corresponding to injection parameter values (I) 0.0148 and 0.0226 are chosen from the experimental study of Mathur and Dutton [2-3] as test cases and have been investigated in details. For both the cases, the supersonic freestream over the cylindrical afterbody had a Mach number 2.46, the static pressure of 28700Pa and stagnation temperature of 300K leading to a flow with velocity 574 m/s. The exit velocity of subsonic bleeder nozzle corresponding to a particular injection parameter is found out from the given velocity profile in the measurements [2]. The pressure and temperature are given by the following Equations (2) & (3).

Adiabatic wall with no slip boundary condition has been applied on the cylindrical afterbody. Supersonic free-stream boundary condition with pressure, velocity and temperature same as the supersonic inlet has been applied on the outer boundary of the cylindrical computation domain, while at the outlet a non-reflecting convective outflow condition is used [27-32]. For LES simulation, the physical time step or $\Delta t$ is kept $1\times10^{-6}$s corresponds to the highest value of CFL number (in the region of refinement) in order of 0.6, whereas the value of CFL drops rapidly as we move further from the base surface. Simulations are carried out for several flow-through times, while the time averaging of the flow field is achieved over ~40+ flow-through times to obtain better statistics.

## 2.2. Meshing Strategy and Computation Domain

In this study, dimensions of geometry and the boundary conditions are selected according to the experiments performed by Mathur and Dutton [2-3]. According to their experiments, we have selected the radius of cylindrical afterbody ($R_o$) to be 31.75mm while the radius of the coaxial bleed nozzle opening on the base surface is 12.7mm. Figure 1(a) shows the schematic of computation domain and Figure 1(b) shows the strategy adopted for meshing the computation domain. To obtain proper domain size, three different computation domain sizes are initially studied. One is having a radial span of $5R_o$ & axial span of $10R_o$, and other two grids have a similar axial span of $20R_o$ and different radial span of $10R_o$ & $15R_o$. As expected like the previously published literature [27], all the grids produce similar results and the error lies within 2%. Hence, the grid with a radial span of $10R_o$ & axial span of $20R_o$ is considered for further study along the similar line with the published literature, while the approach length has been considered to be $4R_o$ (as previously used [27]) for both the bleed ratios (I=0.0148 & I=0.0226). Since the region of interest is the near wake of the cylindrical afterbody, the fine mesh has been used near the base and walls. Further downstream and towards radially outside the domain, the mesh has been stretched to optimize the computation cost. Since we are using a structured solver, O-H type blocking strategy is employed in the region behind the based to reduce the skewness and to achieve a better quality of mesh so that the convergence issues can be dealt with at a higher CFL number.

Figure 1 illustrates the dimensions of the computation domain and mesh preparation strategy for the mesh with approach length $4R_0$. To study the grid independence, tests are performed on three different meshes with $4R_0$ approach length. One mesh has 1.43 million grid points and the other two have 1.85 million and 2.66 million grid points. The finer meshes have been designed with more points in the near wake of the base along the axial direction so that the dynamics of flow structures in the shear layer bounded recirculation region can be resolved properly, whereas the resolution in the radial direction and azimuthal direction remains same in all three grids. In Figure 2(a), it can be easily seen that the distribution of normalized axial velocity, kinetic energy, and Reynolds-stress obtained from medium and fine meshes (1.85M & 2.66 M grid points) match closely with each other. Furthermore, it is observed that the results obtained from different meshes are in good agreement with each other in the near wake region while a little discrepancy can be observed in the location away from the base. The result obtained from 1.43 million mesh is not comparable with the results obtained from other two meshes, possibly owing to the fact that the coarsest mesh has excessively stretched grid in the far wake region which leads to inadequate mesh resolution. In addition to that, we have reported the %Error ($\left(U_{compt} - U_{exp}\right) \times 100 / U_{exp}$) for centerline axial velocity which shows that the error lies within 5% except between $1.5 < x/R_o < 2.7$. Since there is no experimental data available neither for kinetic energy nor Reynolds-stress components, similar error analysis can not be carried out. However, these components are compared in the base-flow case without control and reported in previously published work [27], and this grid is the extension/modification to that framework only. Therefore, for further study the mesh with 1.85 million grid points is used for the geometry with approach length $4R_0$ as used in the previously published work [27].

Initially, the calculations are carried out for two different time steps before choosing the final time step. These time steps are calculated based on stability criteria using the smallest grid size in the domain and upstream velocity, although we are using an implicit time discretization scheme. The simulations are performed using two different time steps, i.e. $\Delta t = 3 \times 10^{-7}$s & $\Delta t = 1 \times 10^{-6}$s, which are well below the stability limits. Fig. 2(b) shows the residual ($L_2$ norm) and Maximum-residual during the each physical iteration while using 20 sub-iterations for each physical iteration. As it turns out that 10 sub-iteration is good enough to get converged solution in a physical time step. Moreover, solutions using both the time steps fall below the specified residual level before reaching 20 sub-iterations. One thing to be noted here is that the plotted residual (L$_2$ norm) and Maximum-residual are the markers of worst cell in the computational domain, means rest of the elements/cells lie well below these limits.

From Figure 3(a), it can be discerned that the log-log plot of resolved energy spectrum obtained from the mesh with 1.85 million grid points seems to approach -5/3 slope in the inertial subrange and thus the grid can be considered to be fine enough to resolve the large eddy structures in the wake region. Moreover, using the method proposed by Celik et al. [43] which uses eddy viscosity ratio as an indicator for examining the grid resolution. The formulation of this method is given by

$$LES\_IQ = \frac{1}{1 + 0.05 \left( \frac{\upsilon_{t,eff}}{\upsilon} \right)^{0.53}} \quad [24]$$

In this formulation $\upsilon_{t,eff}$ denotes the effective viscosity, which is the sum of laminar and turbulent viscosity, and $\upsilon$ denotes the laminar viscosity. The LES quality index should be greater than 0.8 for good LES predictions [44]. Figure 3(b) depicts the LES quality index distribution throughout the domain and it confirms the quality of the present grid. Since the finer mesh has better resolution in the wake region, it is chosen for the rest of the calculations. Henceforth, we have reported the detailed results using this mesh (1.85 million grid points) only.

## 3. RESULTS & DISCUSSION

### 3.1. Salient Features of the flow field

Schematic diagram of the salient axisymmetric flow features of supersonic base flow and effects of the introduction of base bleed have been shown in Figure 4. In the case of supersonic base flow without mass bleed (Fig. 4(a)) Prandtl-Mayer expansion waves are formed at the base corner as the flow turns and the incoming turbulent boundary layer detaches at the base corner forming a free shear layer. Further downstream, the free shear layer rapidly expands and eventually experiences adverse pressure gradient as the flow realigns itself forming a reattachment shock system. The free shear layer separates the low-velocity recirculation region from the supersonic flow outside. The free shear layer has a dominating role in the complicated fluid dynamics of recirculation region as intensive turbulent mixing and energy transfer occur across the shear layer. Fluid in the recirculation region is entrained and accelerated towards the reattachment zone by the shear layer and towards the reattachment, it encounters adverse pressure gradient. However, the fluid in the recirculation region has very low momentum. As the fluid fails to overcome the adverse pressure gradient, it reaches a stagnation point from where it is sent back towards the base. The fluid sent back from the reattachment zone encounters another stagnation point near the base and the Primary Recirculation Region (PRR), depicted in Fig. 4(a), is formed. Now when a subsonic jet is injected into the recirculation region, momentum is added to the fluid in the recirculation region and low velocity becomes capable of penetrating further through the adverse pressure gradient in the reattachment zone; thus both the stagnation points are pushed downstream along with the PRR (Fig. 4(b)). In the process, the wake region widens and the turning angle of the supersonic flow at the base corner also reduces. Quantitative measurement of the displacement of PRR will depend on the momentum of the fluid being injected. Another noticeable flow feature that appears due to injection of the subsonic jet in the wake region, is the formation of secondary recirculation regions (SRR) between the subsonic jet stream and the outer free shear layer. As the subsonic jet is injected into the recirculation region containing low-velocity fluid, an inner shear layer is formed between the jet and the low-velocity fluid in the recirculation region. This inner shear layer entrains the low-velocity fluid in recirculation region towards the PRR where it encounters the first stagnation point of PRR (closer to the base) and an adverse pressure gradient, the fluid particles with lower momentum are sent back towards the base. A portion of the low momentum fluid which forms the SRR closer to the jet is again entrained by the outer shear layer, and some of it encountering the adverse pressure gradient at the downstream is again sent back towards the base and thereby forming another SRR closer to the outer shear layer. Thus, a considerable portion of the fluid entrained by the outer shear layer comes from the bleed. Moreover, due to the presence of the SRRs, the fluid with relatively higher momentum originating from bleeder shields the base surface from the shear layer and the low-velocity fluid trapped in PRR leads to a higher pressure distribution on the base surface with respect to base flow with no mass bleed.

As the injection parameter is increased, the momentum flux of the fluid in the bleeder also increases and the fluid in near wake penetrates further through the adverse pressure gradient. Thus the PRR is pushed further downstream and it becomes smaller and on the other hand, the sizes of the SRRs increase as well (Fig. 4(c)). If the injection parameter is increased monotonically, a situation would arise when the fluid in the recirculation region would have enough momentum to penetrate through the adverse pressure gradient of reattachment zone and this would also lead to much more developed SRRs with more fluid trapped in the SRRs. So with increasing value of injection parameter, initially the pressure on the base surface increases as the SRRs increase in size, but with a further increment of injection parameter beyond an optimal value, the SRRs become large enough to impose detrimental effects on base pressure.

### 3.2 Mean flow properties

In this study, we have performed an analysis based on two different injection parameters ($I$), respectively, 0.0148 and 0.0226. Figure 5 shows the velocity profile of the subsonic jet at the bleed nozzle. It can be seen that, with our imposed boundary condition, the velocity profiles obtained from LES studies match with the experimental data for both values of $I$ . Figure 6 exhibits the distribution of mean axial velocity along the center line and mean pressure on the base surface for $I = 0.0148$ and $I = 0.0226$ along with the flow without control. From these plots it can be inferred that the LES calculation exhibits reasobaly good match with the experimental measurements for both the bleed ratios.

It is to be noted that the base bleed stabilizes the near wake by delaying the interaction of top and bottom shear layers in the wake and hence suppresses the vortex shedding. The attenuation of vortex strength in the near wake region leads to an increase in the base pressure, resulting in drag reduction. The approach length allows the turbulent boundary layer (TBL) to grow on the base wall before it is separated at the base corner. As the TBL grows, the higher velocity gradient along wall normal direction also increases. Thus after separation, the free-shear layer will have a higher shear strain, and this might be the cause for higher turbulence production in the shear layer leading to a higher turbulent mixing rate. This corroborates the afore-mentioned explanation for the over-prediction of velocity as depicted in Fig. 6.

The contour plots (Fig. 7) of mean pressure obtained from the cases with injection parameter values of 0.0148 and 0.0226 strongly suggest the presence of expansion wave at the base corner and reattachment shock system down-stream of the wake; however, the qualitative comparison of the pressure contour reveals the occurrence of a wider and delayed recompression region in the wake region for the case with higher injection. Velocity contours with streamlines presented in Figure 7 exhibit that a considerable portion of fluid entrapped within SRRs originated from the bleeder jet and a considerable amount of fluid entrained by the outer shear layer also comes from the bleeder jet. If the plots in Figure 7 are compared, it can be discerned that SRRs, in the case of I=0.0226, are more developed than the ones appearing in the case of I=0.0148. With I=0.0148, the subsonic jet seems to decelerate till the point of flow reversal while in the case with I=0.0226, flow reversal does not occur though the jet stream decelerates. The higher velocity injection leads to a wider wake as expected.

Furthermore, it has also been observed that the core of the recirculation region (Fig. 6) is shifted when compared to experiments for both the cases studied herein. The shifting appears to be similar as reported in the literature [27] and the possible reasons are: (a) the difference in boundary layer thickness close to the base corner with the measurements. As the computational boundary layer thickness is thicker compared to the experimental one, the amount of higher momentum fluid, which enters in the domain, is also more in the computations. Therefore, the growth of the shear layers is affected and the core has been shifted downstream. (b) Secondly, the eddy viscosity predictions may affect the growth of the shear layers as reported in the literature [27]. Hence, the discrepancies may come due to either of these possible reasons or due to both of them; it is a bit difficult to identify one over other and that can be considered as a part of future work which requires detailed parametric investigation.

## 3.2. Unsteady Flow Structures

For finding out the periodicity of the shedding of vortical structures from the outer shear layer, data is probed in the outer shear layer to obtain time-series pressure data and consecutively power spectral density (PSD) analysis has been performed. For both the injection parameters, PSD has revealed the peak at a frequency of 4.8kHz (Fig. 8). Further, the study of the evolution of flow structures over one cycle of the observed frequency has been carried out by plotting a sequence of 5 phase-averaged contour plots of vorticity magnitude at Z=0 plane along with streamlines, uniformly spanning over one cycle of $360^0$, for both values of I. Figure 9 shows the phase-averaged plots at five different time instances representing phases during a complete cycle of the observed oscillation frequency (4.8 kHz), and the data at each phase is averaged over ten of these cycles in order to eliminate some of the turbulent fluctuations arising from the higher frequency oscillations. From the plots, it is evident that the vortices are generated in the outer shear layer due to shear layer instabilities and they gradually travel downstream along the shear layer and grow in size before they detach from the shear layer at the recompression region. In phase 1(Fig. 9), newly formed vortices in the outer shear layer close to the base can be observed. These vortices seem to have grown and moved downstream in phase 2. In phase 3, while these vortices seem to have moved further downstream, new vortices start to appear in the upstream locations. In phase 5, the older vortices seem to get detached from the outer shear layer. This periodical shedding of vortices is also associated with the flapping of the shear layer. Evolutions of vortices which are generated in the inner shear layer (between subsonic jet and fluid trapped in SRRs) have been observed as well. Vortices generated in the inner shear layer due to K-H instabilities seem to grow as they travel downstream and eventually detach from the shear layer, meanwhile another vortex starts to form in the upstream region. To better understand these flow structures, POD analysis is performed and discussed in the later subsection.

To further quantify the above-mentioned flow structures, we have reported the instantaneous vorticity contours for different flow conditions with and without control. Figure 10 represents a comparison between instantaneous vorticity contours at Z=0 plane for injection parameter values of 0.0148, 0.0226 and supersonic base flow without mass injection. For both the cases with mass-injection, it can be observed that the larger structures with higher vorticity appear in the reattachment zone, as depicted in the phase-averaged plots (Fig. 9). Near the reattachment zone, the vortical structures seem to rapidly detach from the shear layer and then they grow as we move further downstream. Similar structures emanating from the shear layer has been observed in the earlier study of supersonic baseflow without mass injection (Fig. 10(c) )[27]. In the sequences of the vorticity contours, it has been noticed that hairpin-like vortical structures are being issued from the shear layer near the reattachment zone. It has also been observed that periodical shedding of vortical structure is associated with the flapping of the shear layer. In the recompression region, presence of the recompression shock system and high-temperature gradient has led to a high contribution from the baroclinic term $\left(\frac{\vec{\nabla}\rho \times \vec{\nabla} p}{\rho^2}\right)$(Fig. 11) in the vorticity transport equation, which might be the reason for the formation of convoluted vortical structures with high vorticity magnitude. It is also noteworthy to mention that in case of the higher value of injection parameter, the shear layer inclination angle is less and the wake is much wider compared to the lower injection parameter, whereas lower inclination of the shear leads to weaker reattachment shock system. The second shear layer formed around the jet seems to have dominating effects on the dynamics of the flow structures inside the recirculation region.

Notably, the shedding of vortices issued due to Kelvin-Helmholtz instability can also be clearly observed around the subsonic jet.

To identify the 3-D coherent structures in the flow, iso-surfaces of Q-criterion colored with vorticity magnitude have been plotted in Figure 12. Mathematical formulation of Q-criterion is given by equation (4).

$$Q = -\frac{1}{2}(S_{ij}S_{ij} - \Omega_{ij}\Omega_{ij})$$ [4]

where $\Omega_{ij}$ is the rotation rate tensor and $S_{ij}$ is the strain rate tensor. Figures 12(a) depict the flow structures in the outer shear layer for I=0.0148 and I=0.0226 respectively. It can be clearly seen that smaller structures lie inside the recirculation region while larger structures with higher vorticity magnitude appear in the outer shear layer. Azimuthal coherence between the structures in the Q-criterion plot has not been observed. Yet it can be noticed that there are hairpin-like 3-D vortical structures with high vorticity magnitude, which are being issued from the shear layer and seem to grow as they move in the downstream direction and eventually detach from the shear layer near the reattachment zone. Finally, at the further downstream behind the recompression region, these structures seem to form larger coherent structures in the far wake region. Vortical structures with similar shape are also seen in the sequence of vorticity magnitude plot and it has also been noticed that there is a sudden periodicity in the formation of these structures. Figures 12(b) also exhibit the flow structures in the inner shear layer for I=0.0148 and I=0.0226. In that case also, similarities between the flow structures can be observed for both the injection parameters. In addition, ring-like structures originated from K-H instability near the jet inlet surrounding the subsonic jet can also be observed in both the cases. It is also worth noticing that the structures near subsonic jet inlet have higher vorticity magnitude in the case of higher injection parameter, as expected. Furthermore, to justify these structures POD analysis is performed and discussed in the following sub-section.

### 3.3. POD analysis

For the POD analysis, data is extracted from a rectangular region at Z=0 plane, i.e. along the mid-plane. After that, the data is extrapolated on the 2D uniform grid while considering a large number of snapshots, i.e. 100. Three different grid resolutions($100 \times 100, 200 \times 200$ $and$ $300 \times 300$) have been considered to study the effect of grid resolution on the POD results. Three different time-steps (time difference between two consecutive snap-shots) respectively $3 \times 10^{-5}$ $seconds, 6 \times 10^{-5}$ $seconds$ $and$ $9 \times 10^{-5}$ $seconds$ are also considered to study the effects of varying time-step on the POD results. The study of grid variation and time-step variation are performed for the injection parameter value I=0.0148 only considering the fact that both the injection parameters under study have shown almost similar flow structures and the time-scale of the evolving structures in the outer shear layer is also found to be the same for both the cases. Grid convergence study using both the energy and enstropy based POD analysis (Fig. 13) shows that for all three grid resolutions, the distribution of eigen values over eigen modes have similar pattern while the values converging towards the higher resolution. The time-step variation study (Fig. 13) also shows that the eigen values tend to converge as larger time-step is used. From the distributions of normalized eigen values obtained from the grid convergence and time-step convergence it has been concluded that the grid resolution of $300 \times 300$ and the time-step of $9 \times 10^{-5}s$ is adequate for conducting further study.

Figure 14(a) shows the comparison between distributions of eigen values across eigen modes obtained from energy based POD analysis for both the injection parameters. Figure 14(b) exhibits a similar comparison for the enstropy based POD analysis. It is clear from these figures that the basic shape of the distribution of eigen values is similar to both the values of injection parameters. Since enstropy is more related to the first order derivative of the fluctuating velocity field, the basis functions forming the enstropy field are different from the basis functions forming the velocity field. Thus it is expected that there are going to be some differences in the POD modes obtained from energy based POD and enstropy based POD. If the normalized eigen values obtained from energy based POD and enstropy based POD are compared, it can be seen that the distribution of normalized eigen values is different as tabulated below while comparing with the base flow case without control.

| Modes | Normalized Eigen Value (base flow case without mass bleed) | |
|---|---|---|
| | Energy | Enstropy |
| 1st | 0.068 | 0.045 |
| 2nd | 0.044 | 0.033 |
| 3rd | 0.041 | 0.027 |
| 4th | 0.032 | 0.023 |
| 5th | 0.029 | 0.019 |

Table 1: Normalized eigen value for base flow without mass bleed using both energy and enstropy based POD [27]

| Modes | Normalized Eigen Value (I=0.0148) | | Normalized Eigen Value (I=0.0226) | |
|---|---|---|---|---|
| | Energy | Enstropy | Energy | Enstropy |
| 1st | 0.058 | 0.025 | 0.049 | 0.027 |
| 2nd | 0.045 | 0.024 | 0.042 | 0.023 |
| 3rd | 0.040 | 0.022 | 0.036 | 0.020 |
| 4th | 0.032 | 0.021 | 0.032 | 0.019 |
| 5th | 0.029 | 0.020 | 0.027 | 0.017 |
| 6th | 0.025 | 0.019 | 0.026 | 0.019 |

Table 2: Normalized eigen value for two bleed ratios using both energy and enstropy based POD

Left column of Figures 15-17 exhibit the first few modes obtained from energy based POD for the cases with base flow (without mass bleed [27]), I=0.0148 and I=0.0226, respectively, where the contour indicated the distribution of fluctuating component of energy (obtained from first three components of the POD modes which correspond to the fluctuating velocity components) across the domain and the vectors are formed from the in-plane components of fluctuating velocity field (length of the vectors shown are not in proportion with the magnitude of the in-plane components of fluctuating velocity field). The right column of Figures 16 &17 exhibit the first six enstropy modes of the cases I=0.0148 and I=0.0226 respectively over the domain. As the POD modes have been arranged in descending order of their corresponding eigen values, the POD modes appearing earlier have more contribution towards the flow field and the structures shown in those modes are more likely to appear. The contour plots of first five POD modes obtained from energy and enstropy based POD do not exhibit exact similar mode distributions across the plane; however, it is noteworthy to mention that there are some intrinsic similarities. The first mode of energy based POD of I=0.0148 (Fig. 16) case shows the presence of high energy concentration in the outer shear layer and the recompression region, while the presence of vortical structures can be observed near the inner shear layer. On the other hand, the first mode of enstropy based POD of I=0.0148 case exhibits the presence of distinct structures with a concentration of vorticity magnitude in the outer shear layer and near the inner shear layer suggesting the presence of vortical structures in those regions. The second and third modes of both energy and enstropy based POD of the I=0.0148 case suggest the presence of distinct vortical structures in the outer shear layer.

While in the second mode the structures in the outer shear layer seem to appear close to the recompression region, the third mode exhibits the presence of this structure in the outer shear layer closer to the base. The vectors presented in the energy based POD also suggest the presence of vortical structures in the inner shear layer. From the contours presented in the energy and enstropy based POD modes, it can be concluded that these vortical structures present in the inner shear layer contain less energy than the ones present in the outer shear layer. Now if the energy and enstropy based POD modes obtained from the case with I=0.0226 are considered, it can be seen that they also suggest the presence of similar kind of flow structures in the outer shear layer. But it can be noted that the first energy based POD mode of the case with I=0.0148 suggests reattachment of the outer shear layer; while in the case of I=0.0226, none of the POD modes suggest reattachment, instead periodical shedding of larger vortical structures can be observed in the first energy based POD mode. It is evident from this study that the most energetic flow structures remain in the outer shear layer while the vector plots obtained from energy based POD also suggest violent mixing of the fluid coming from bleeder in the near wake. Periodical shedding of vortical structures from the outer shear layer have also been observed and this is in accord with what is observed earlier (presence of hairpin-like structures) in the vorticity magnitude plots and iso-surface plots of Q-criterion (Fig. 12). The flow structures revealed by the POD modes support our earlier comment about the evolution of vortices generated due to instability in the outer shear layer which is observed through phase averaging in the previous section. If we compare the dominant POD modes obtained from the current numerical study of base flow with mass injection, with the POD modes of simple base flow case without any flow control (Fig. 15), it can be observed that the wake region widens due to the mass injection. Comparison between the $1^{st}$ POD mode of all three cases suggests that mass bleed increases the turbulent kinetic energy concentration in the shear layer and higher amount of turbulence in the shear layer leads to a higher rate of momentum transfer from the outer flow to the wake region leading to significant pressure recovery on the base surface.

## 4. CONCLUSIONS

The evolution of flow structures in supersonic base flow with mass bleed is studied numerically using LES. LES has predicted the mean axial velocity distribution along the center line and the mean pressure distribution over the base surface with acceptable accuracy. Unsteady flow structures such as large eddies in the downstream region, smaller vortices inside the recirculation region, hairpin-like structures near the reattachment region are also properly captured by the LES. Azimuthal coherence in the flow structures appearing in the outer shear layer has not been observed in any of the cases, yet periodical shedding of hairpin-like vortical structures from the outer shear layer near the recompression region has been observed. The Q-criterion plots clearly show the presence of this hairpin-like vortical structures in the recompression region. The flapping of the outer shear layer has also been observed and that has to be associated with the shedding of vortical structures, while the frequency of flapping of the outer shear layer has found to be same for both the injection parameters. Phase averaged streamline plots exhibit the evolution of vortices in the outer shear layer. Apparently, the vortical structures which are generated in the outer shear layer due to instabilities rapidly grow as they move downstream and eventually detach from the shear layer near the reattachment zone. Shedding of vortical structures from the inner shear layer has also been observed. Later, when energy and enstropy based POD analysis is performed, the POD modes also confirm the presence of vortical structures with concentrated energy and enstropy in the outer shear layer.

**Acknowledgments**

Simulations are carried out on the computers provided by the Indian Institute of Technology Kanpur (IITK) (www.iitk.ac.in/cc) and the manuscript preparation as well as data analysis has been carried out using the resources available at IITK. This support is gratefully acknowledged.

(a)

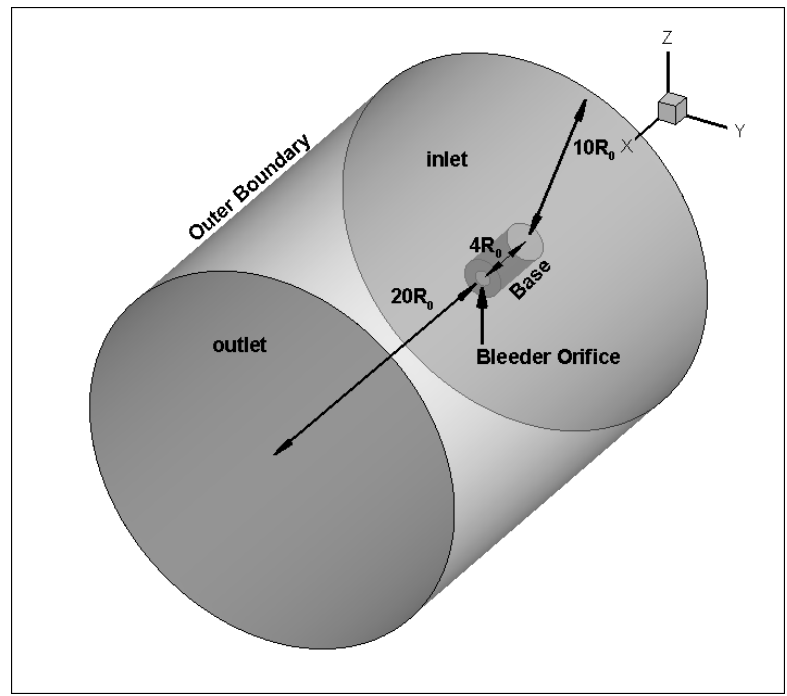

(b)

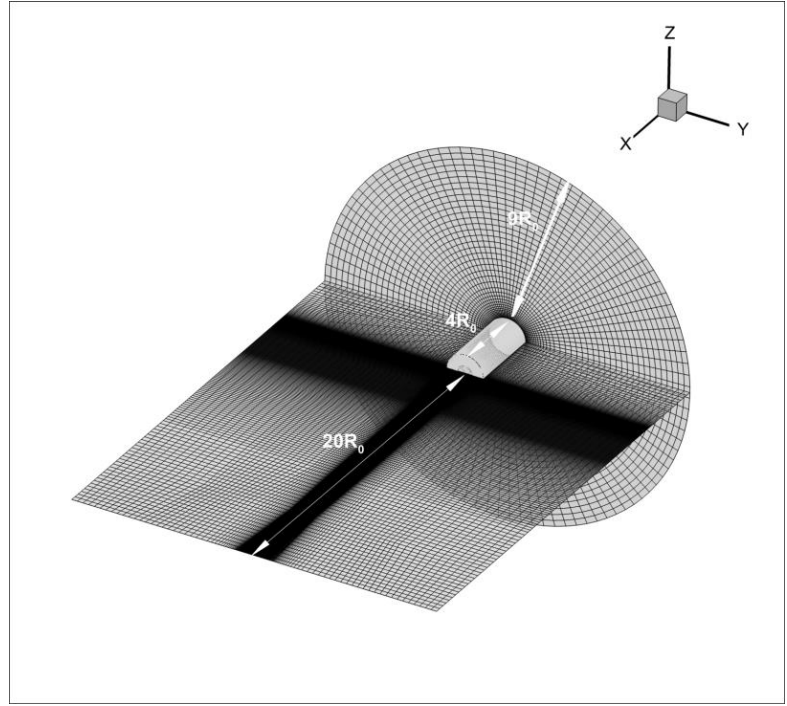

Figure 1: Illustration of mesh preparation strategy

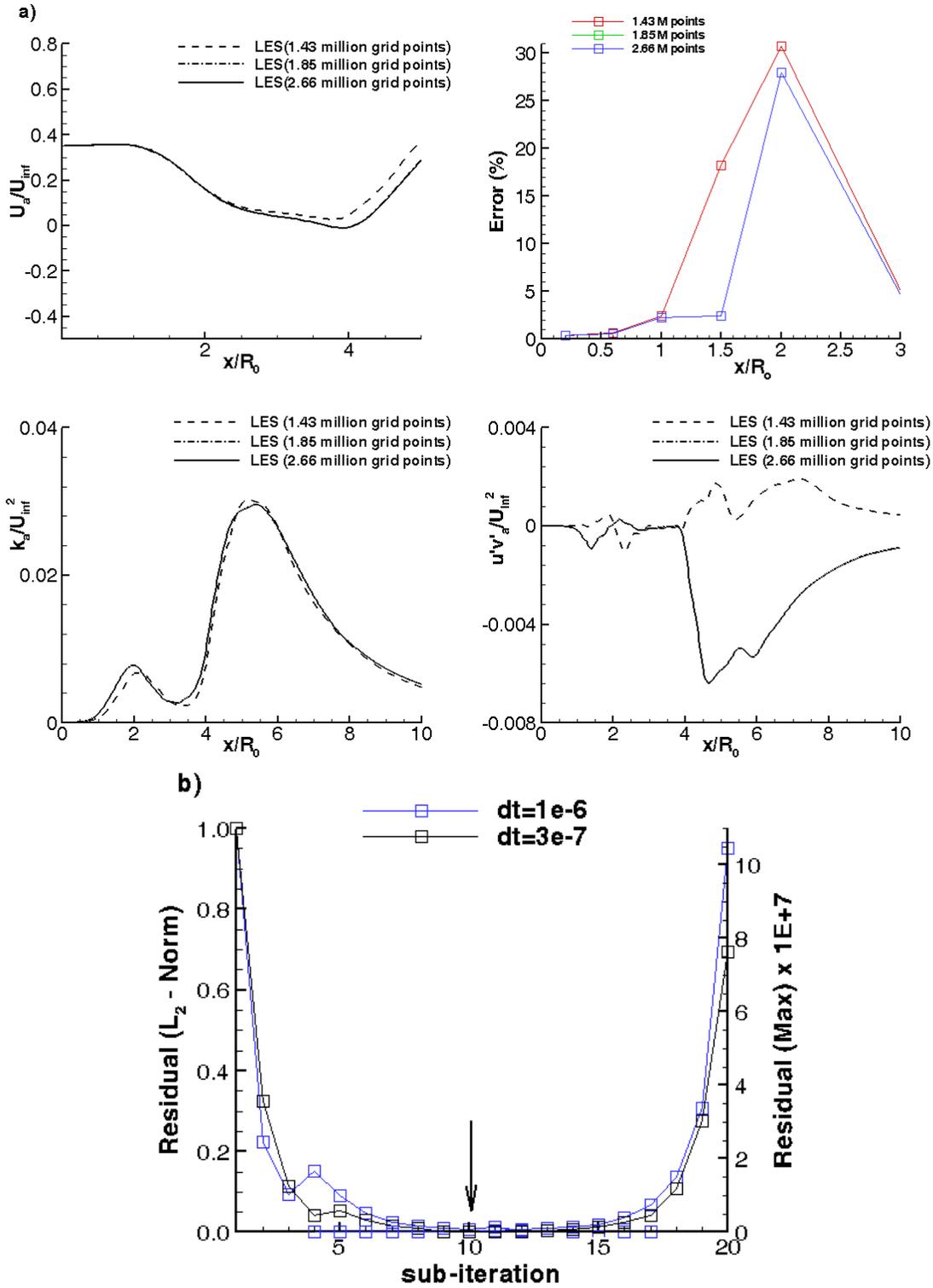

Figure 2: Results for (a) grid independence (b) time independence

(a)

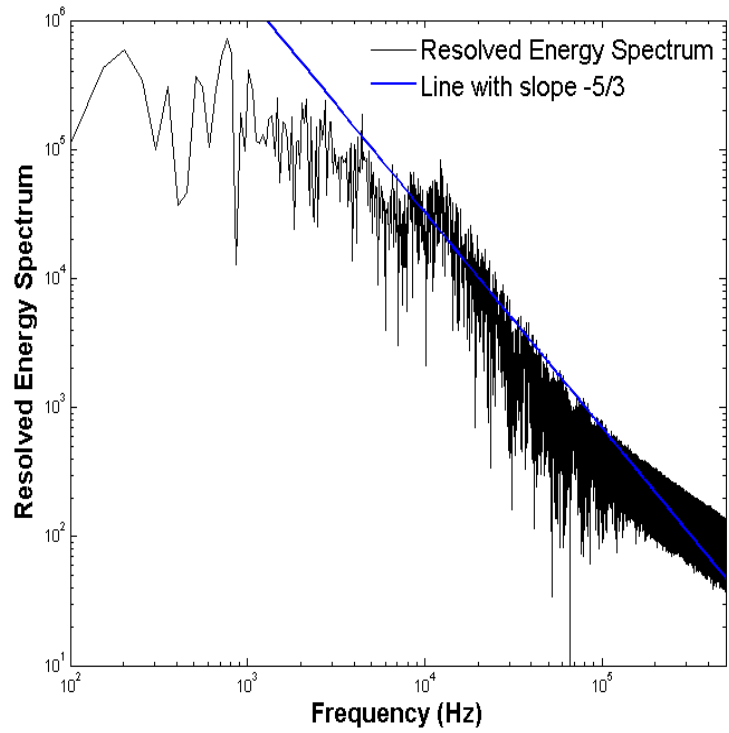

(b)

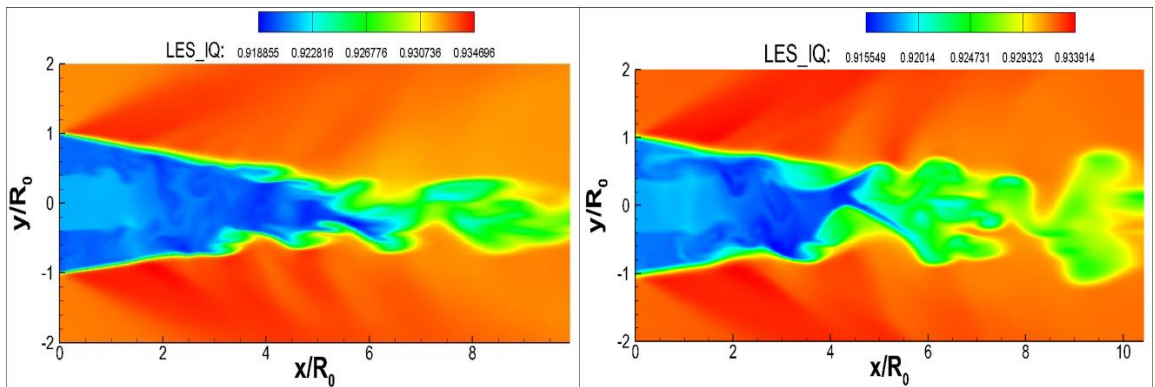

Figure 3: (a) Resolved Energy Spectrum at a point in recirculation region obtained from the mesh with 1.85 million grid points, (b) Resolution of Grid using LES quality criteria [43] for I=0.0148(left) & I=0.0226 (right)

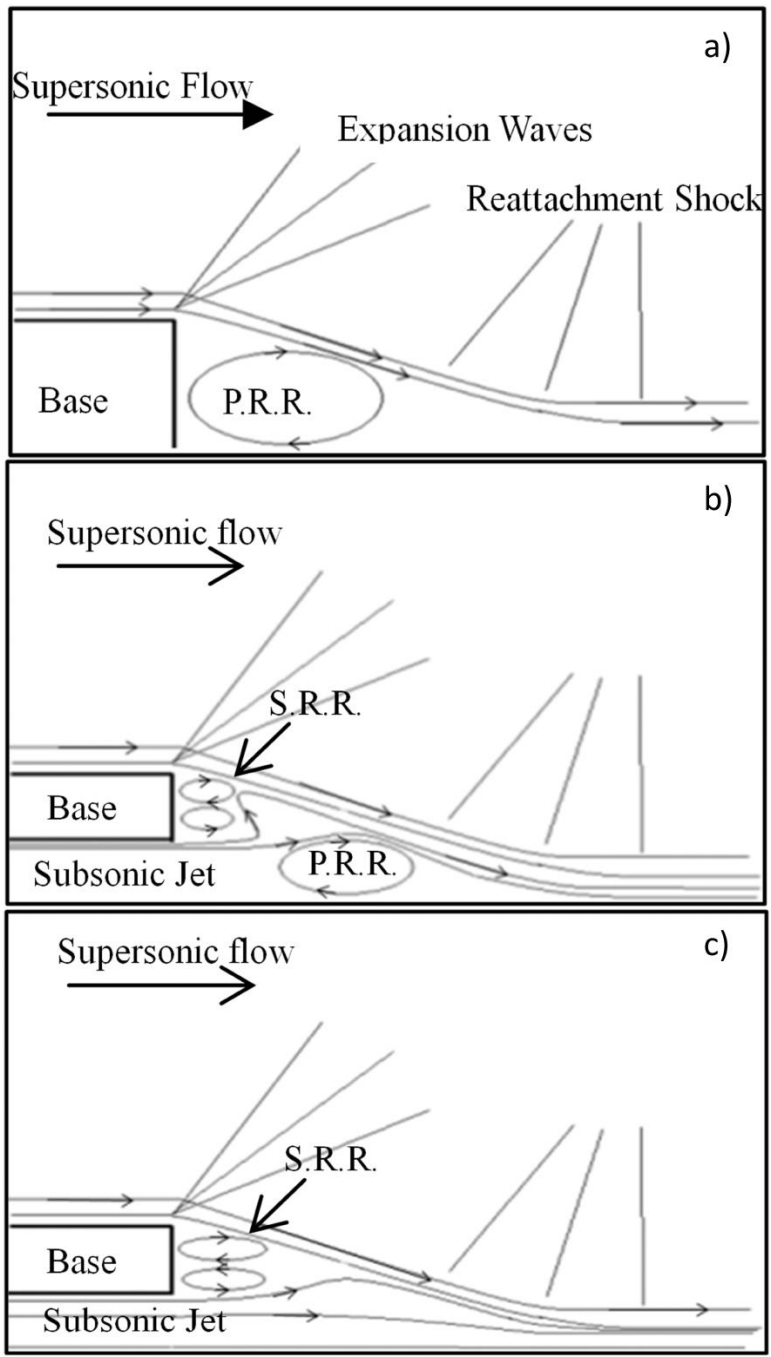

Figure 4: (a) Supersonic Base Flow, (b) Supersonic Base Flow with low mass bleed rate, (c) Supersonic Base Flow with high mass bleed rate

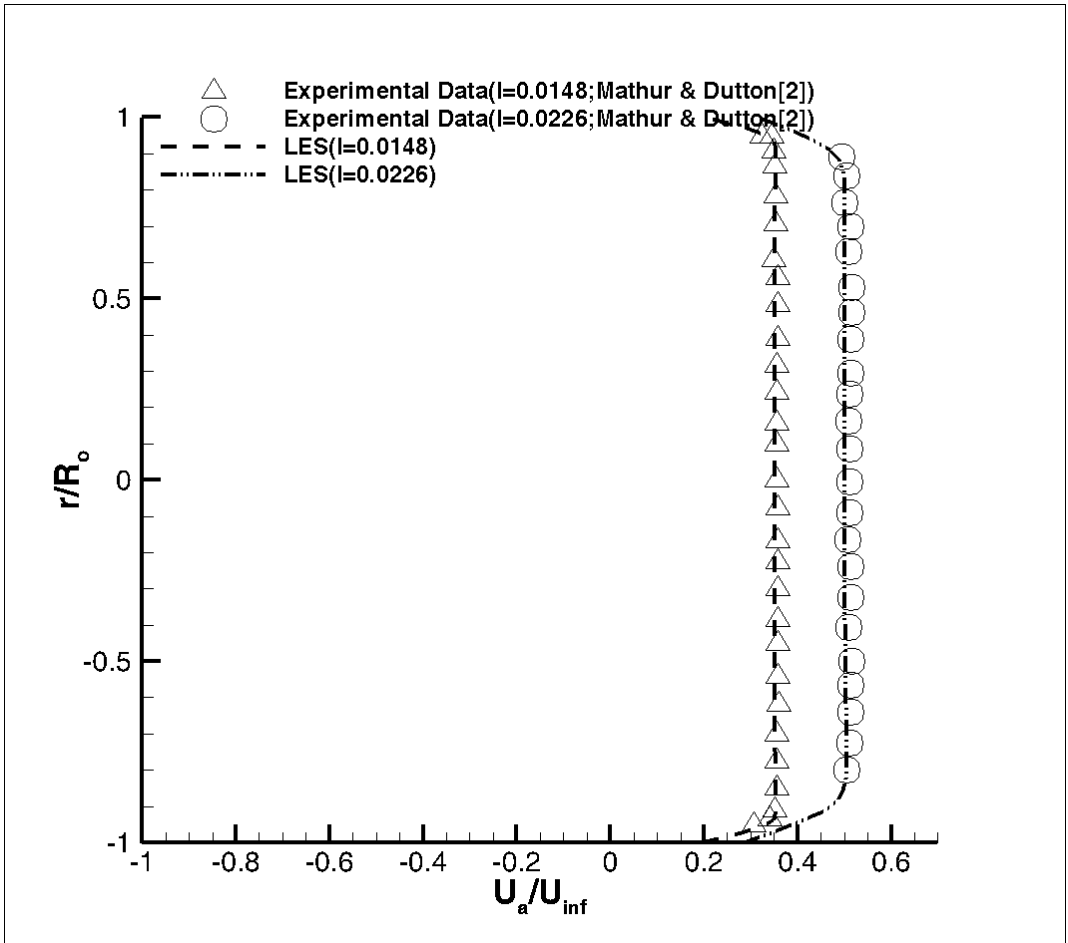

Figure 5: Jet Velocity Profile

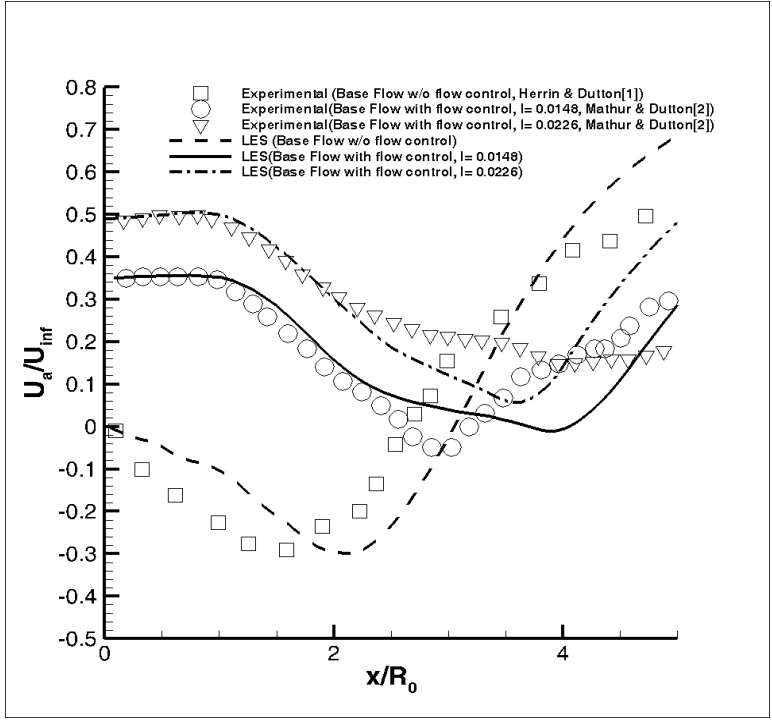

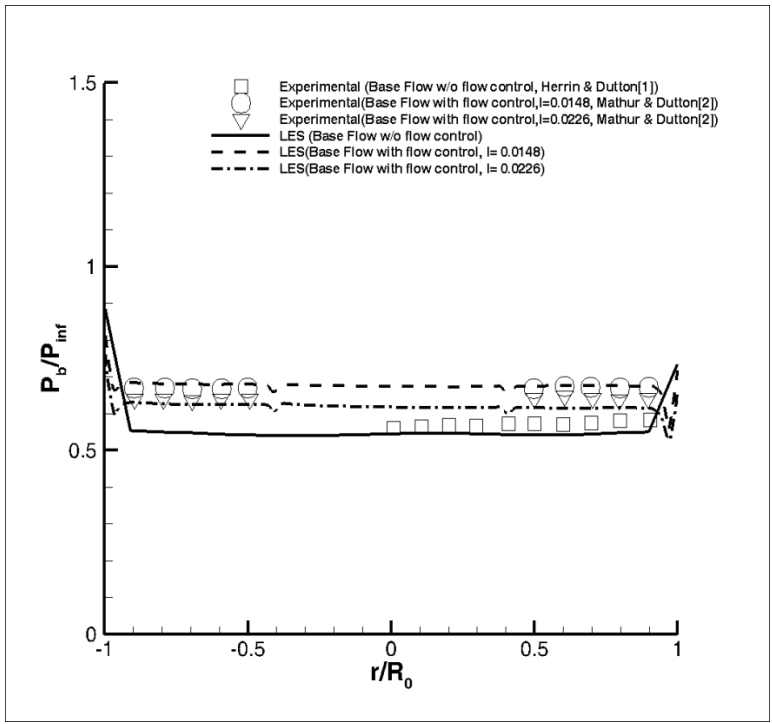

Figure 6: Distribution of Mean axial velocity along center-line and mean pressure on base-surface along diameter for base flow with and without control

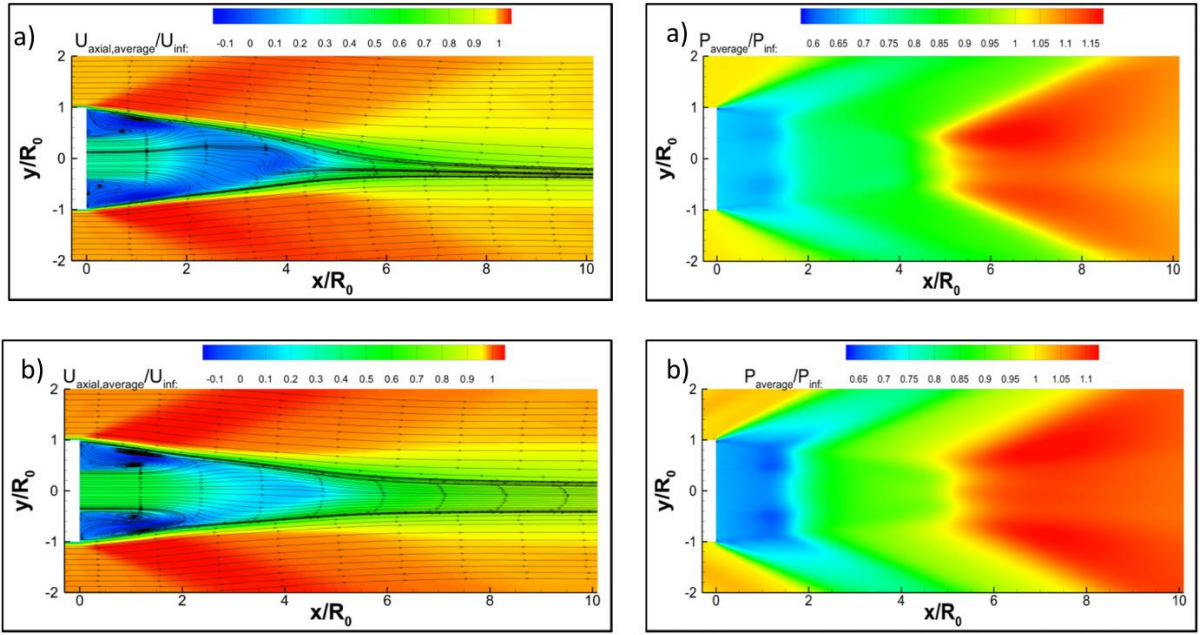

Figure 7: Mean axial velocity and Pressure contours: (a) I=0.0148 (top), (b) I=0.0226 (bottom)

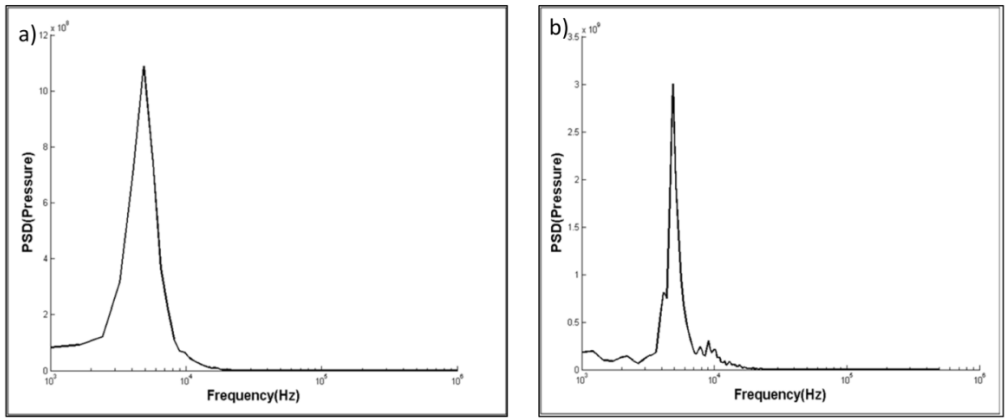

Figure 8: PSD of pressure at a point in shear layer for: (a) I=0.0148, (b) I=0.0226

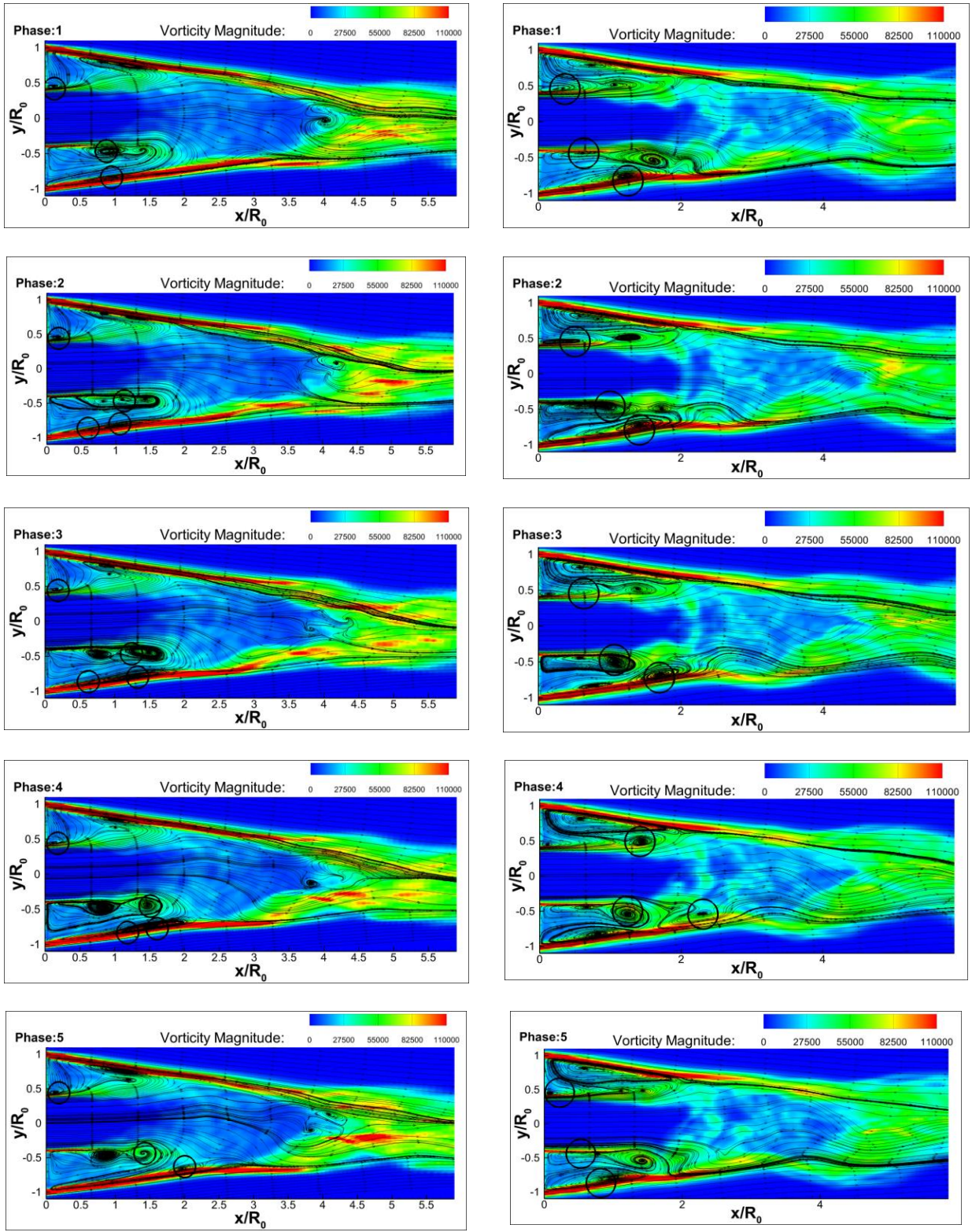

Figure 9: Phase averaged vorticity contour along with streamlines for I=0.0148(left column) and I=0.0226 (right column)

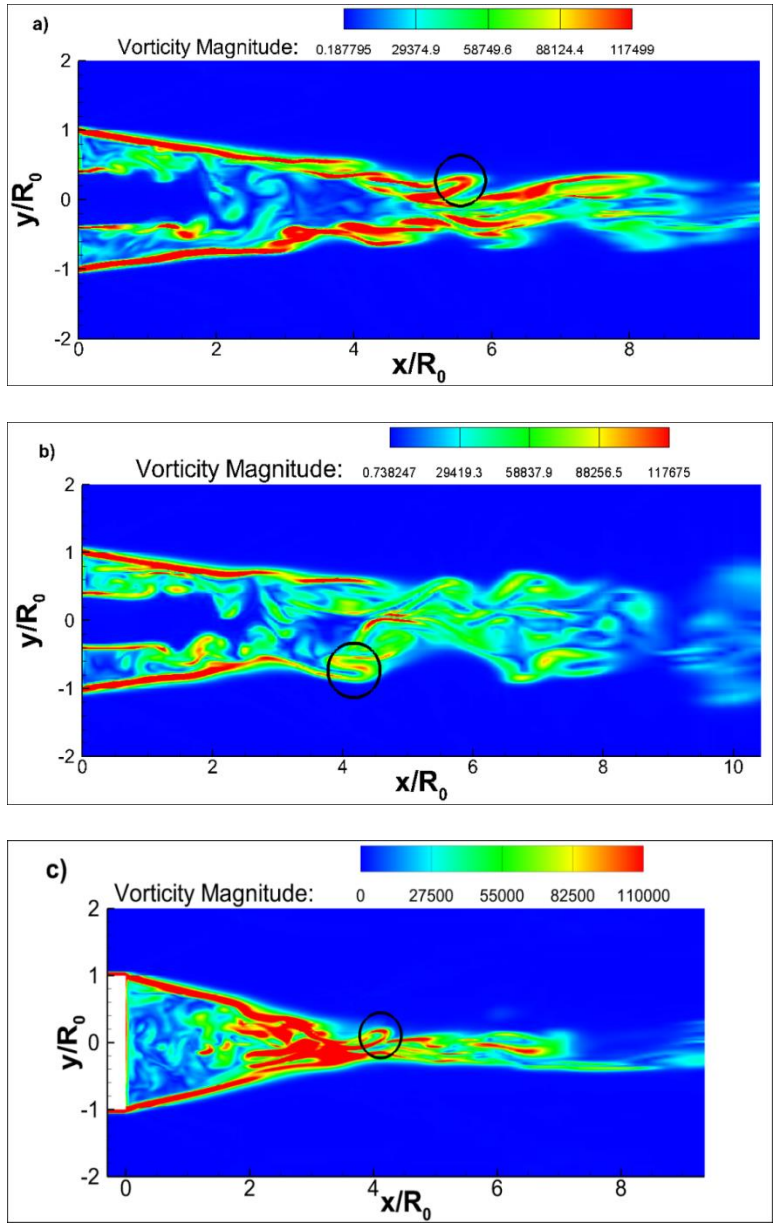

Figure 10: Instantaneous vorticity contour at Z=0 plane for: (a) I=0.0148, (b) I=0.0226 and (c) I=0

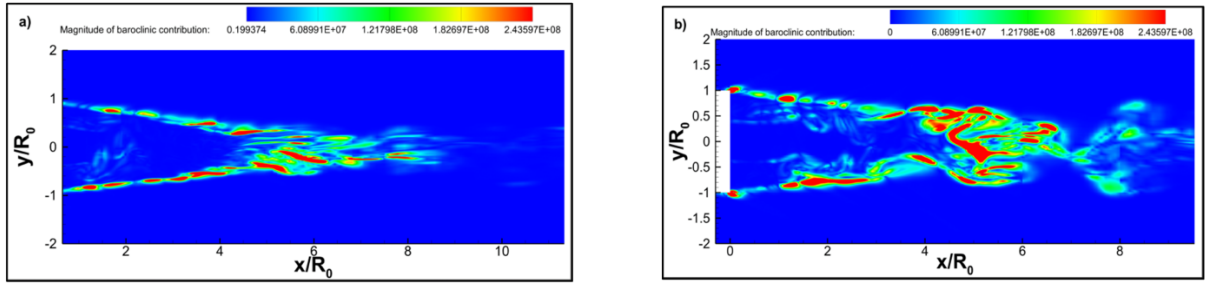

Figure 11: Contour plot of magnitude of baroclinic contribution term at Z=0 plane for: (a) I=0.0148 and (b) I=0.0226

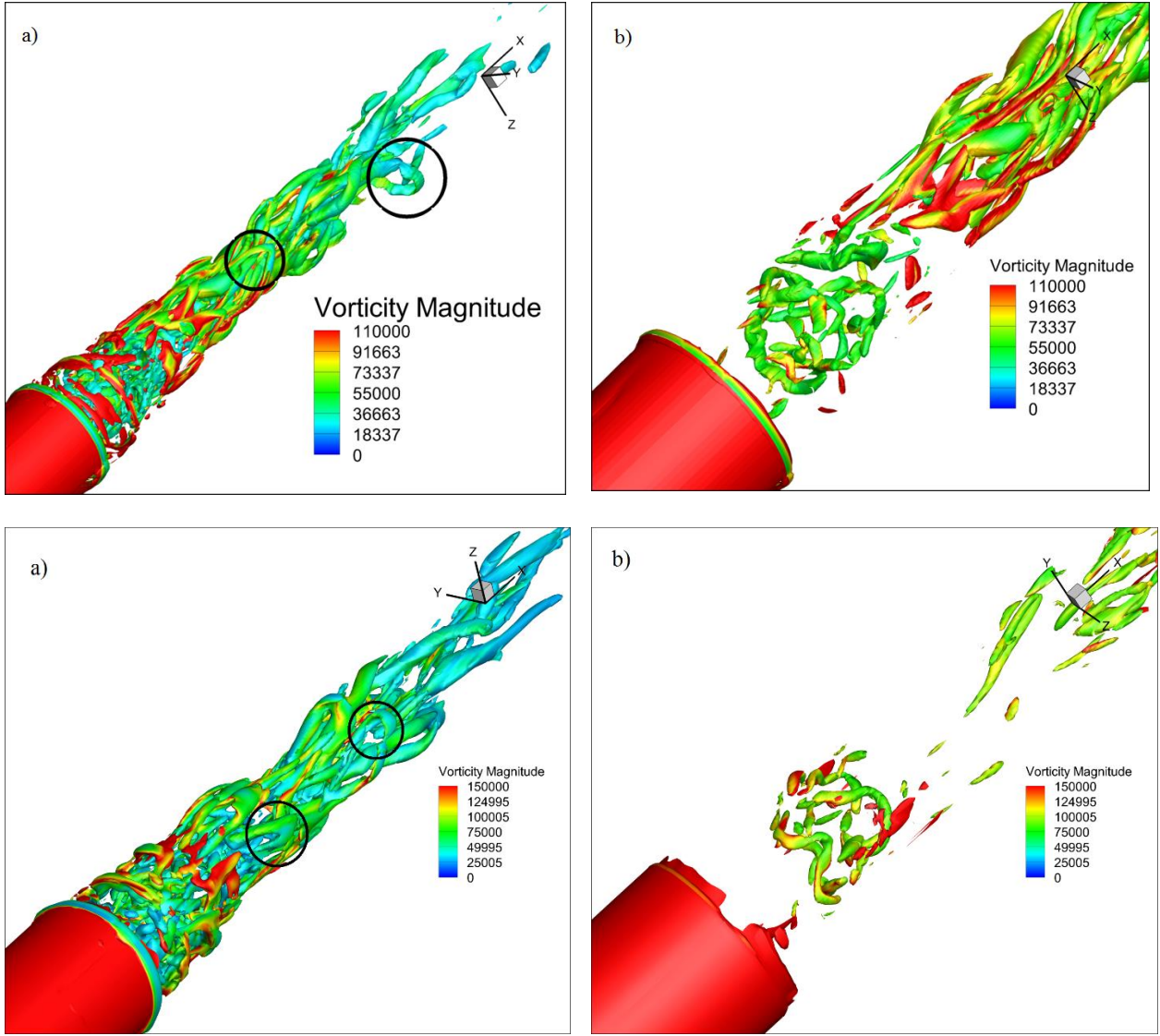

Figure 12: Iso-surface plot of Q-criterion showing the flow structures for I=0.0148 (top) and I=0.0226 (bottom): (a) outer shear layer, (b) inner shear layer

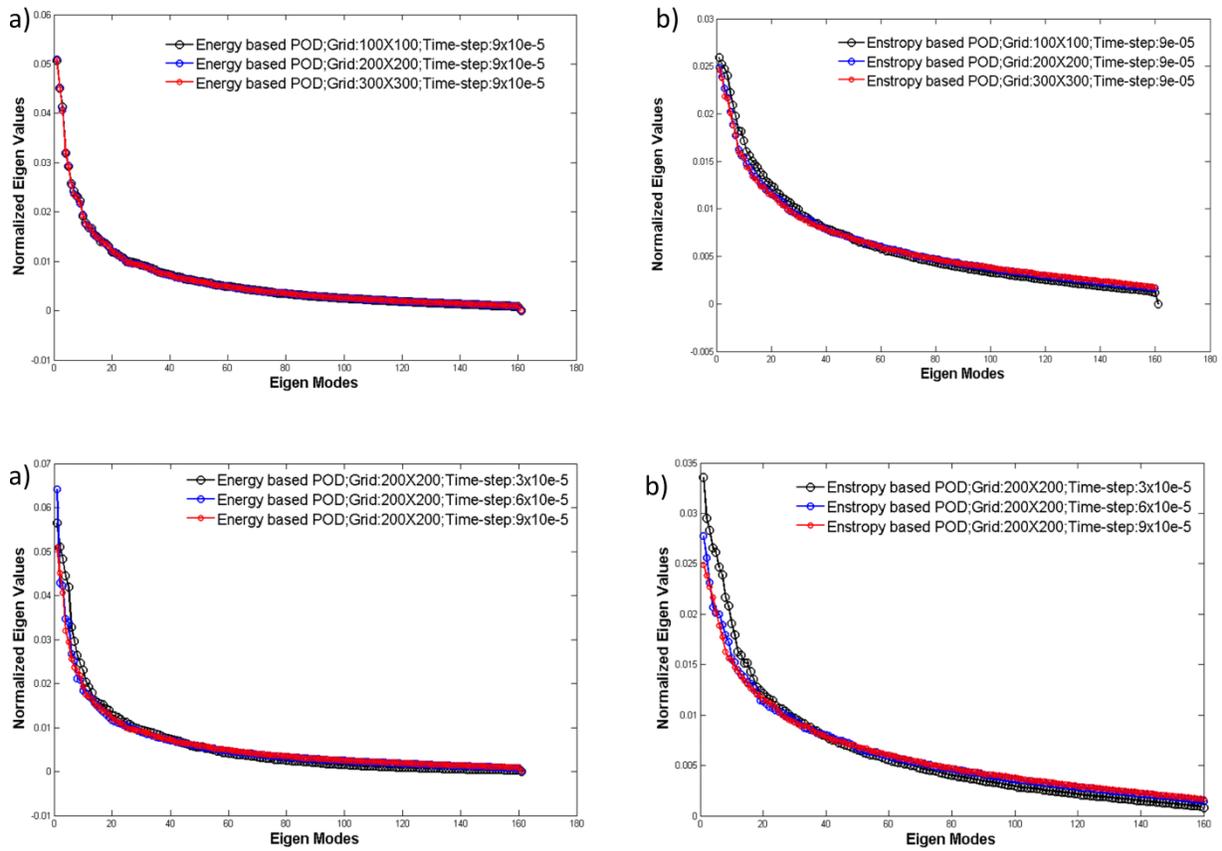

Figure 13: Grid and time-step convergence study using energy and enstropy based POD for I=0.0148

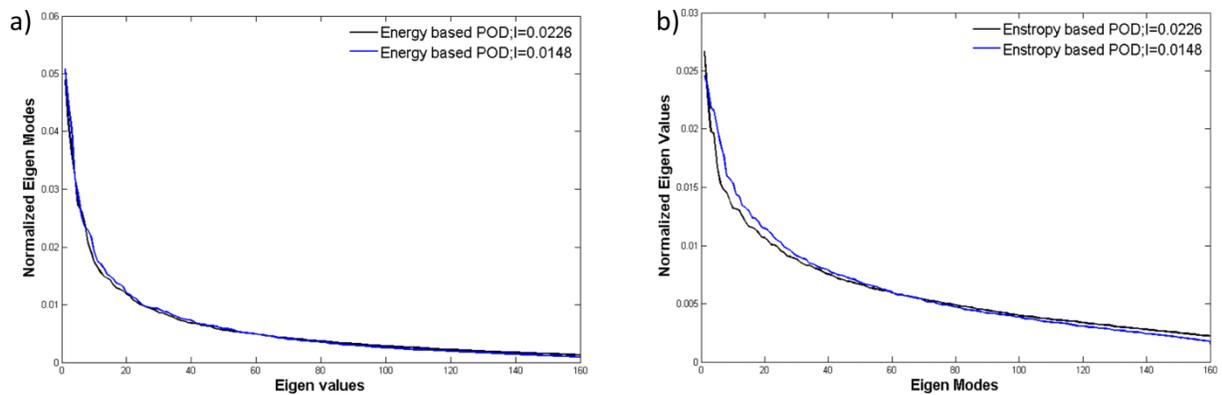

Figure 14: Eigen value distribution obtained from energy and enstropy based POD

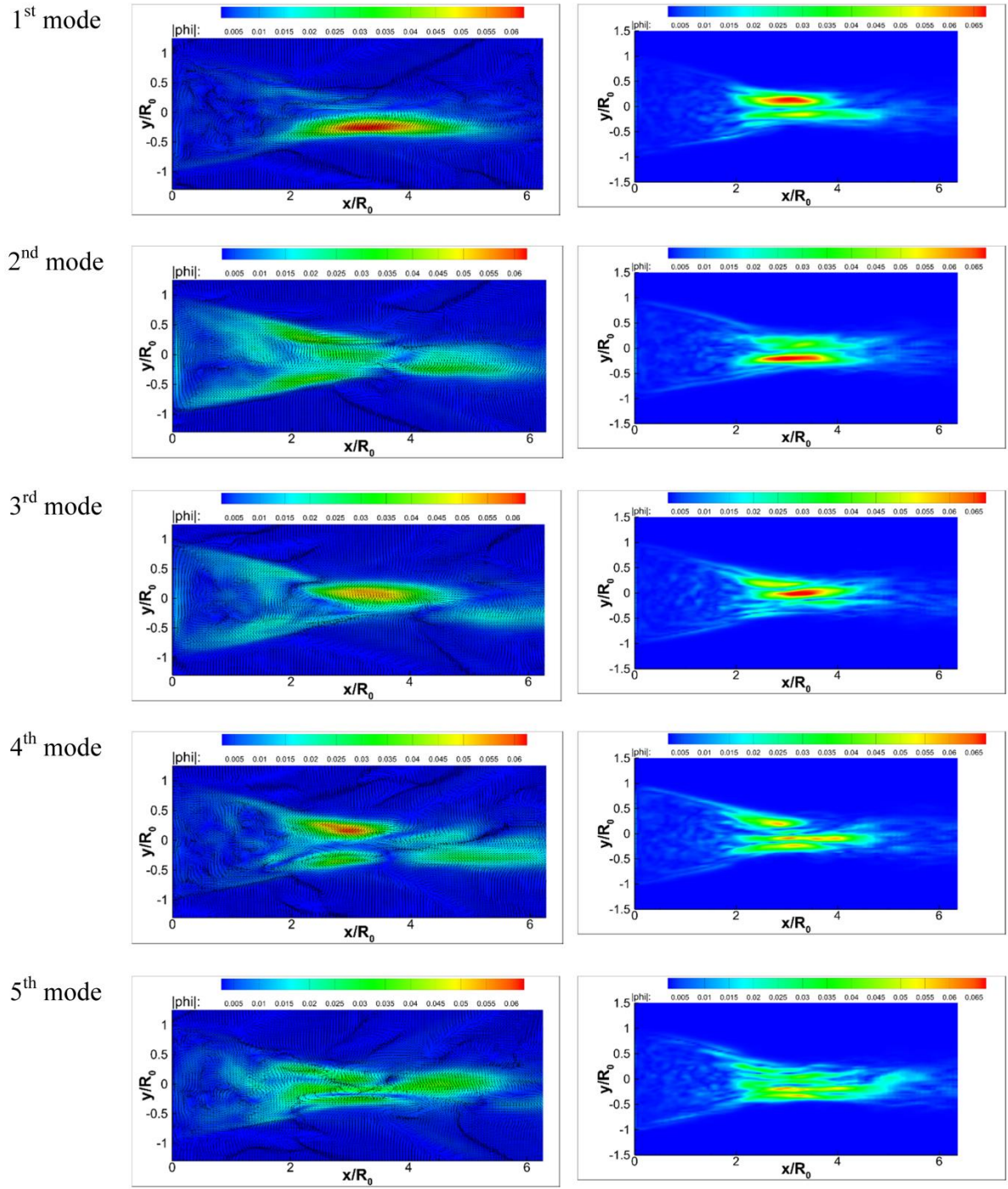

Figure 15: Different modes obtained from both energy based (left column) and enstropy based (right column) POD for Supersonic base flow without mass bleed [27]

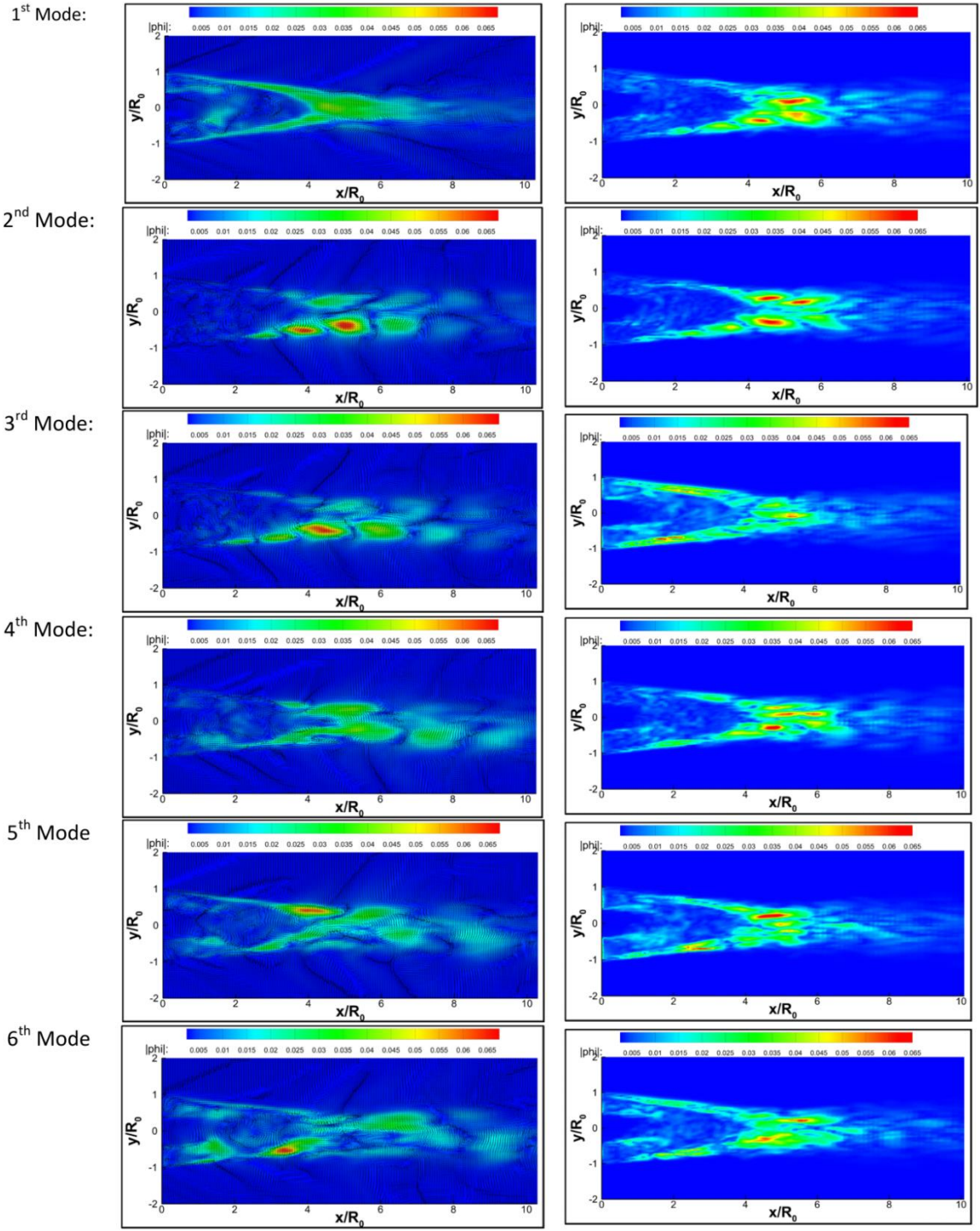

Figure 16: Different modes obtained from both energy based (left column) and enstropy based (right column) POD for the injection parameter I=0.0148

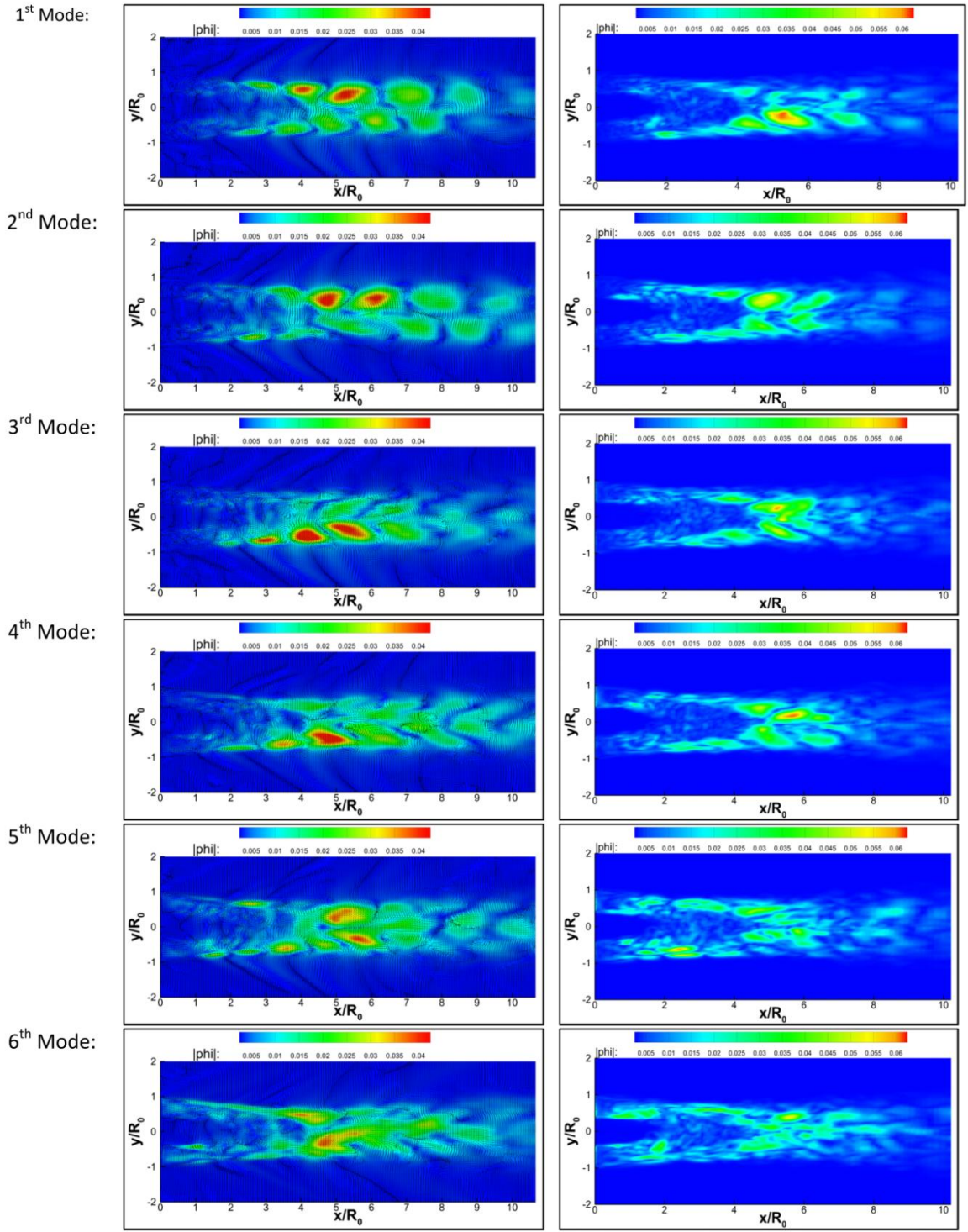

Figure 17: Different modes obtained from both energy based (left column) and enstropy based (right column) POD for the injection parameter I=0.0226